\documentclass[pra,aps,amsmath,amssymb,twocolumn]{revtex4}
\newcommand{\hh}{{\mathcal{H}}}
\newcommand{\iu}{{\mathtt{i}}}

\newcommand{\xdif}{{\mathrm{d}}}
\newcommand{\Tr}{{\mathrm{Tr}}}
\newcommand{\spc}{{\mathrm{spec}}}

\newcommand{\bro}{\boldsymbol{\rho}}
\newcommand{\vbro}{\boldsymbol{\varrho}}

\newcommand{\bsg}{\boldsymbol{\sigma}}
\newcommand{\pen}{\openone}

\newcommand{\lasf}{{\mathsf{\Lambda}}}
\newcommand{\am}{{\mathsf{A}}}
\newcommand{\bn}{{\mathsf{B}}}
\newcommand{\km}{{\mathsf{K}}}
\newcommand{\elm}{{\mathsf{L}}}
\newcommand{\mn}{{\mathsf{M}}}
\newcommand{\nm}{{\mathsf{N}}}
\newcommand{\ax}{{\mathsf{X}}}
\newcommand{\ay}{{\mathsf{Y}}}
\newcommand{\pq}{{\mathsf{p}}}
\newcommand{\qp}{{\mathsf{q}}}

\begin{document}
\clearpage
\preprint{}

\title{Entropic uncertainty relations for successive measurements of canonically conjugate observables}

\author{Alexey E. Rastegin}
\affiliation{Department of Theoretical Physics, Irkutsk State University,
Gagarin Bv. 20, Irkutsk 664003, Russia}

\begin{abstract}
Uncertainties in successive measurements of general canonically
conjugate variables are examined. Such operators are approached
within a limiting procedure of the Pegg--Barnett type. Dealing
with unbounded observables, we should take into account a
finiteness of detector resolution. An appropriate reformulation of
two scenarios of successive measurements is proposed and
motivated. Uncertainties are characterized by means of generalized
entropies of both the R\'{e}nyi and Tsallis types. The R\'{e}nyi
and Tsallis formulations of uncertainty relations are obtained for
both the scenarios of successive measurements of canonically
conjugate operators. Entropic uncertainty relations for the
case of position and momentum are separately discussed.
\end{abstract}

\maketitle

\pagenumbering{arabic}
\setcounter{page}{1}

\section{Introduction}\label{sec1}

The Heisenberg uncertainty principle \cite{heisenberg} has been
recognized as one of the fundamental scientific concepts. The
Heisenberg's thought experiment was first examined rather
qualitatively. An explicit formal derivation was firstly given in
\cite{kennard}: the product of the standard deviations of position
and momentum in the same state cannot be less than $\hbar/2$. To
arbitrary pair of observables, this approach was extended by
Robertson \cite{robert}. This very traditional formulation has
been criticized \cite{deutsch,maass}. It gives no
characterization, when a prepared state commutes with any of the
two observables. The first entropic uncertainty relation for
position and momentum was derived by Hirschman from the
Hausdorff--Young inequality \cite{hirs}. His result has later been
improved in \cite{beck,birula1}. The improved relation was used in
derivation of entanglement criteria for continuous variables
\cite{walborn09}. Entropic relations in multi-dimensional position
and momentum spaces were obtained in \cite{huang11}. In the
discrete case, entropic approach yields a non-trivial restriction
whenever the two observables do not share any common eigenvector
\cite{maass}. Entropic uncertainty relations are currently the
subject of active research (see the reviews
\cite{ww10,brud11,cbtw15} and references therein). Questions of
their optimality are addressed in \cite{asmf15}. Other approaches
are based on the sum of variances \cite{huang12,mpati14} and on
majorization relations \cite{prz13,fgg13,rpz14,lrud15,arkz16}. The
variance-based formulation was recently applied to noise and
disturbance \cite{msp16}. Measure-independent notions of joint
uncertainty of several variables are studied in \cite{gour16}.

The most traditional formulation known as preparation uncertainty
relations assumes repeated trials with the same state. For each of
the chosen measurements, the corresponding probability
distribution is taken. Using some quantitative figures of
uncertainty, we then study bounds on the total measure of
uncertainty in the scenario. Of course, other measurement
scenarios could be considered here. Some of them are related to
the case of successive measurements \cite{mdsrin03,paban13}. At
each stage, an actual pre-measurement state somehow depends on the
results of previous measurements. In a certain sense, this
situation is more realistic in the context of quantum information
processing. Uncertainty relations are now interesting not only
from the conceptual viewpoint. These studies are stimulated by a
recent progress in using quantum systems as an informational
resource \cite{renner10,renner11,nbw12}. Today, physicists are
able to carry out experiments with individual quantum systems
\cite{wineland13,haroche13}. In quantum information processing,
our subsequent manipulations usually deal with an output of the
latter stage. The Heisenberg's thought experiment with microscope
should rather be treated as a scenario of measurement uncertainty
in successive measurements \cite{blw13}. Apparently, studies of
quantum uncertainties in the scenarios of successive measurements
have received less attention than they deserve \cite{paban13}. The
authors of \cite{paban13} also discussed links between
uncertainties in successive measurements and Ozawa's treatment of
noise and disturbance \cite{ozawa03,ozawa04}.

For the scenarios of successive measurements, entropic uncertainty
relations were considered only in finite-dimensional settings
\cite{mdsrin03,bfs2014}. For a pair of qubit observables, this
approach was further developed with the use of R\'{e}nyi's
entropies \cite{zzhang14} and Tsallis' entropies \cite{rastctp15}.
In the present work, we will deal with successive measurements
that are not projective. The paper is organized as follows. In
Section \ref{sec2}, we review the required material concerning
entropic functions and application of the Pegg--Barnett formalism
to canonical conjugacy. In Section \ref{sec3}, we discuss the two
scenarios of successive measurements of canonically conjugate
operators. This reformulation is necessary since we consider
observables with continuous spectra. Entropic measures of
uncertainty also differ somehow from the measures used in the
discrete case. In Section \ref{sec4}, we derive R\'{e}nyi and
Tsallis formulations of uncertainty relations for both the
scenarios of successive measurements of canonically conjugate
operators. The position-momentum case is separately addressed. In
Section \ref{sec5}, we conclude the paper with a brief summary of
results obtained.

\section{Preliminaries}\label{sec2}

In this section, we recall the required material and fix the
notation. Uncertainties will be characterized via the R\'{e}nyi
and Tsallis entropies, which have many convenient properties
\cite{bengtsson}. Then we consider necessary details of the
Pegg--Barnett formalism.

Let us consider distributions with a discrete label. For the given
probability distribution $\pq=\{p_{i}\}$, its R\'{e}nyi
$\alpha$-entropy is defined as \cite{renyi61}
\begin{equation}
R_{\alpha}(\pq):=\frac{1}{1-\alpha}{\>}\ln\left(\sum\nolimits_{i} p_{i}^{\alpha}\right)
 , \label{rpdf}
\end{equation}
where $0<\alpha\neq1$. For $0<\alpha<1$, the R\'{e}nyi
$\alpha$-entropy is a concave function of the probability
distribution. For $\alpha>1$, it is neither purely convex nor
purely concave \cite{jizba}. In the limit $\alpha\to1$, the
formula (\ref{rpdf}) gives the standard Shannon entropy
\begin{equation}
H_{1}(\pq)=-\sum\nolimits_{i}p_{i}\ln{p}_{i}
\ . \label{shan}
\end{equation}
For the given probability distribution $\pq=\{p_{i}\}$ and any
$0<\alpha\neq1$, the Tsallis $\alpha$-entropy is defined as
\cite{tsallis}
\begin{align}
H_{\alpha}(\pq)&:=\frac{1}{1-\alpha}\,
\left(\sum\nolimits_{i} p_{i}^{\alpha}-1\right)
\nonumber\\
&=
-\sum\nolimits_{i} p_{i}^{\alpha}\,\ln_{\alpha}(p_{i})
\ . \label{tsent}
\end{align}
Here, the $\alpha$-logarithm is given by
\begin{equation}
\ln_{\alpha}(\xi):=\frac{\xi^{1-\alpha}-1}{1-\alpha}
\ . \label{lathdf}
\end{equation}
Note that the $\alpha$-entropy (\ref{tsent}) is concave for all
$0<\alpha\neq1$. For $\alpha\to1$, the $\alpha$-logarithm reduces
to the usual one so that (\ref{tsent}) reduces to (\ref{shan}).

In view of non-additivity, the Tsallis entropy is well known in
non-extensive thermostatistics \cite{tsallis}. Up to a factor, the
$\alpha$-entropy (\ref{tsent}) is actually identical to an
information measure introduced earlier in \cite{havrda}.
Incidentally, entropic functions of this type have found use far
beyond the context of thermostatistics. For instance, such
information measures were applied in formulation of Bell
inequalities \cite{rastann} and in studies of combinatorial
problems \cite{ecount}.

In the following, we will use a convenient notion similar to
vector norms. For $\beta>0$, we define
\begin{equation}
\|\pq\|_{\beta}:=
\left(\sum\nolimits_{i} p_{i}^{\beta}\right)^{1/\beta}
 . \label{betnrm}
\end{equation}
The right-hand side of (\ref{betnrm}) gives a norm only for
$\beta\geq1$. For $\alpha\neq1$, one can rewrite (\ref{rpdf}) in
the form
\begin{equation}
R_{\alpha}(\pq)=\frac{\alpha}{1-\alpha}{\>}\ln\|\pq\|_{\alpha}
\ . \label{rprpdf}
\end{equation}
The $\alpha$-entropy (\ref{tsent}) is expressed in terms of
$\|\pq\|_{\alpha}$ as well.

Let us proceed to the case of continuously changed variables. In
principle, the formulas (\ref{rpdf}) and (\ref{tsent}) can be
rewritten immediately. If the variable is distributed according to
the probability density function $w(y)$, then
\begin{equation}
R_{\alpha}(w):=\frac{1}{1-\alpha}{\ }
\ln\!\left(\int [w(y)]^{\alpha}\xdif{y}\right)
 , \label{recon}
\end{equation}
where $0<\alpha\neq1$. The integral is assumed to be taken over
the interval of values, for which $w(y)$ is defined. As a rule,
such intervals are clear from the context. It will be convenient
to extend the notion (\ref{betnrm}) to the case of probability
density functions. For the given density function $w(y)$ and
$\beta>0$, we define
\begin{equation}
\|w\|_{\beta}:=
\left(
\int [w(y)]^{\beta}\xdif{y}
\right)^{1/\beta}
 . \label{betynrm}
\end{equation}

Using entropies of the form (\ref{recon}), the following
fact should be kept in mind. In general, they may have negative
values. This is a distinction from (\ref{rpdf}) and
(\ref{tsent}). Since $p_{i}\leq1$, for $\alpha>1>\beta$ we easily
obtain
\begin{equation}
\|\pq\|_{\alpha}\leq1\leq\|\pq\|_{\beta}
\ . \label{paqb1}
\end{equation}
For probability density functions, we cannot generally write
relations similarly to (\ref{paqb1}). If $w(y)\leq1$ for all
acceptable values of $y$, then we truly have
\begin{equation}
\|w\|_{\alpha}\leq1\leq\|w\|_{\beta}
\ . \label{wawb1}
\end{equation}
The latter follows from the normalization $\|w\|_{1}=1$
and $\alpha>1>\beta$. It can be exemplified that the
property (\ref{wawb1}) is not valid for density functions with
sufficiently large variations.

To quantify an amount of uncertainty, we rather wish to deal with
positive entropic functions. Here, one of possible approaches is
to use some discretization. An interval of continuously changed
variable can be divided into a set of non-intersecting bins.
Preparation uncertainty relations with binning were derived by
Bia{\l}ynicki-Birula in terms of the Shannon \cite{IBB84} and
R\'{e}nyi entropies \cite{IBB06}. Such relations can be applied to
entanglement \cite{taska13} and steering detection
\cite{dixon13,howell14}. To reach a good exposition, the size of
the bins should be sufficiently small in comparison with a scale of
considerable changes of the distribution. Let the
bins be described by the set of marks $\{\ell_{i}\}$. Accordingly,
we have the intervals $\Delta\ell_{i}=\ell_{i+1}-\ell_{i}$ with
the maximum $\Delta\ell=\max\Delta\ell_{i}$. We introduce
probabilities
\begin{equation}
p_{i}^{(\ell)}:=\int\nolimits_{\ell_{i}}^{\ell_{i+1}}
w(y){\,}\xdif{y}
\ , \label{qkwx}
\end{equation}
which form the distribution $\pq_{\varDelta}^{(\ell)}$. For this
probability distribution, we can then calculate the entropies
(\ref{rpdf}) and (\ref{tsent}). For sufficiently small bins, these
entropies will provide a good characteristic of uncertainty.

To obtain entropic relations with binning, we will use the
inequalities following from theorem 192 of the book by Hardy et al. \cite{hardy}.
For $\alpha>1>\beta$, we have \cite{barper15}
\begin{align}
&\Delta\ell^{1-\alpha}\,\bigl\|\pq_{\varDelta}^{(\ell)}\bigr\|_{\alpha}^{\alpha}
\leq\|w\|_{\alpha}^{\alpha}
\ , \label{cnv1}\\
&\|w\|_{\beta}^{\beta}
\leq\Delta\ell^{1-\beta}\,\bigl\|\pq_{\varDelta}^{(\ell)}\bigr\|_{\beta}^{\beta}
\ . \label{cnv2}
\end{align}
R\'{e}nyi's entropies of $\pq_{\varDelta}^{(\ell)}$ are generally
unbounded, when $\Delta\ell\to0$ and the other parameters remain
fixed. It is seen from (\ref{cnv1}) and (\ref{cnv2}). In this
limit, we will deal with entropies of continuous distributions
such as (\ref{recon}).

To describe canonically conjugate operators, the Pegg--Barnett
formalism \cite{PBepl,PB89,BP89} will be used. This approach was
originally proposed to fit a Hermitian phase operator. Since
Dirac's famous work \cite{dirac27} on quantum electrodynamics had
appeared, the quantum phase problem has extensively been studied
\cite{lynch}. Dirac's treatment of a quantum phase operator was
later found to suffer from a lot of difficulties \cite{CN68}. A
renewed interest in the quantum phase problem was stimulated by a
progress in quantum optics \cite{lynch}. There is a large
literature on many aspects of this problem \cite{lynch,CN68}.

Instead of infinite dimensions, the Pegg--Barnett formalism begins
with a finite but arbitrarily large state space \cite{PBepl,BP89}.
The final part of the procedure is to find the limit of desired
quantities as the dimensionality tends to infinity. Furthermore,
the concept of canonical conjugacy can be explained from this
viewpoint \cite{pvb90}. It gives additional insights into the
nature of canonically conjugate operators in physically relevant
situations. The notion of complementarity in finite dimensions is
shown to be related with complementarity in the sense of canonical
conjugacy. Entropic uncertainty relations for general canonically
conjugate operators were studied in \cite{gvb95} and by another
method in \cite{barper15}.

We consider the two orthonormal bases $\{|n\rangle\}_{n=0}^{d}$
and $\{|m\rangle\}_{m=0}^{d}$ in $(d+1)$-dimensional Hilbert space
$\hh_{d+1}$. The authors of \cite{pvb90} introduced two Hermitian
operators with the same structure:
\begin{align}
\nm_{d+1}&:=\sum_{n=0}^{d} n{\,}|n\rangle\langle{n}|
\ , \label{nsdf}\\
\mn_{d+1}&:=\sum_{m=0}^{d} m{\,}|m\rangle\langle{m}|
\ . \label{msdf}
\end{align}
They are formally similar to the photon number operator and both
have the spectrum $\{0,1,\ldots,d\}$. To build conjugate
observables, the considered orthonormal bases should obey
\begin{equation}
\langle{n}|m\rangle=(d+1)^{-1/2}{\,}\exp\bigl(\iu{h}(m,n)\bigr)
\ . \label{twb}
\end{equation}
The real-valued function $h(m,n)$ is defined as
$h(m,n)=-(m+b_{0})(n+a_{0})\delta$, where \cite{pvb90}
\begin{equation}
\delta=\frac{2\pi}{d+1}
\ . \label{deldf}
\end{equation}
The parameters $a_{0}$ and $b_{0}$ will be specified later. Here,
we deal with a pair of mutually unbiased bases, whose role in
quantum physics was emphasized by Schwinger \cite{schwinger60}.
The above operators generate shifts in each other's eigenstates
\cite{pvb90}:
\begin{align}
\exp\Bigl(-\iu(\nm_{d+1}+a_{0}\pen_{d+1})j\delta\Bigr)|m\rangle
&=|m+j\rangle
\ , \label{nsht}\\
\exp\Bigl(+\iu(\mn_{d+1}+b_{0}\pen_{d+1})k\delta\Bigr)|n\rangle
&=|n+k\rangle
\ . \label{msht}
\end{align}
The relations (\ref{nsht}) and (\ref{msht}) give a hint for the proper
form of the operators with conjugacy. For a real parameter
$\gamma$, we define
\begin{align}
\ax_{d+1}^{(\gamma)}&:=(\nm_{d+1}+a_{0}\pen_{d+1})\gamma
\ , \label{pmdf}\\
\ay_{d+1}^{(\gamma)}&:=(\mn_{d+1}+b_{0}\pen_{d+1})\delta\gamma^{-1}
\ , \label{qmdf}
\end{align}
These operators should become canonically conjugate for
$d\to\infty$. The eigenvalues of the operators (\ref{pmdf}) and
(\ref{qmdf}) are written as $x_{n}=(a_{0}+n)\gamma$ and
$y_{m}=(b_{0}+m)\delta\gamma^{-1}$, respectively. The differences
between successive eigenvalues are $\Delta{x}_{n}=\gamma$ and
$\Delta{y}_{m}=\delta\gamma^{-1}$. The parameters
$a_{0}$, $b_{0}$, and $\gamma$ are functions of the dimensionality
and should be chosen appropriately \cite{pvb90,gvb95}.

To approach canonical conjugacy, the two possible situations
should be kept in mind \cite{pvb90}. In the first one, the
difference between successive eigenvalues of (\ref{pmdf}) does not
vanish as $d\to\infty$. Accordingly, $\gamma(d)$ should tend to a
non-zero finite limit. In the second case, the conjugate
observables both have continuous spectra. We will focus just on
this case, which includes the position-momentum pair. When
$d\to\infty$, the parameter $\gamma(d)$ approaches zero so that
$(d+1)\gamma$ tends to infinity. It might be inversely
proportional to a power of $(d+1)$ which is between, but not
including, zero and unity \cite{pvb90}. To provide infinite
expansion of spectra in both the negative and positive directions,
we could set $a_{0}(d)=b_{0}(d)=-d/2$ \cite{pvb90}.

The commutator of (\ref{qmdf}) and (\ref{pmdf}) is expressed as
\begin{equation}
\bigl[\ay_{d+1}^{(\gamma)},\ax_{d+1}^{(\gamma)}\bigr]=\delta{\,}[\mn_{d+1},\nm_{d+1}]
\ . \label{ayax0}
\end{equation}
To approach canonical conjugacy, certain physical conditions
should be imposed \cite{pvb90}. Here, the first physical condition
implies that one cannot prepare a state of infinite energy or
momentum. Under such restrictions, the right-hand side of
(\ref{ayax0}) reduces to the form \cite{pvb90}
\begin{equation}
\iu\Bigl(\pen_{d+1}-(d+1)|m=0\rangle\langle{m}=0|\Bigr)
\, . \label{phncom}
\end{equation}
The ket $|m=0\rangle$ corresponds to the minimal eigenvalue
$y_{0}=b_{0}\delta\gamma^{-1}$ of $\ay_{d+1}^{(\gamma)}$. Except
for the term with this ket, the right-hand side of (\ref{phncom})
reproduces the standard commutation relation. The units here are
such that $\hbar=1$. The form (\ref{phncom}) is sufficient for the
first case of canonical conjugacy, including the photon
phase-number pair and the angle-angular momentum pair
\cite{pvb90}. To approach the second case, an additional condition
is required. The relation (\ref{phncom}) can finally be reduced to
the form
\begin{equation}
\bigl[\ay_{d+1}^{(\gamma)},\ax_{d+1}^{(\gamma)}\bigr]_{PP}
=\iu\pen_{d+1}
\, . \label{phpcom}
\end{equation}
It implies that the system must be in a state with a finite
extension in space, or, more generally, with finite moments with
respect to (\ref{qmdf}) \cite{pvb90}. The subscripts in
(\ref{phpcom}) emphasize the role of the above physical
conditions. To derive entropic uncertainty relations, we will
focus on norm-like functionals of the corresponding probability
distributions. Connections between them follow from Riesz's
theorem \cite{riesz27} (see also theorem 297 of \cite{hardy}).

\section{On successive measurements of observables in general}\label{sec3}

In this section, the problem of uncertainty in quantum
measurements is considered for two successive measurements.
Successive projective measurements are physically meaningful only
for observables with purely discrete spectra. In this paper, we
focus on observables with continuous spectra. This case should be
treated in essentially different manner \cite{paban13}. As the
case of projective measurements is conceptually simpler, we
briefly recall it.

Let $\am$ be a finite-dimensional observable, and let $\lasf_{a}$
be the projector onto the $a$-th eigenspace of $\am$. For the
pre-measurement state $\bro$, the probability of outcome $a$ is
equal to $\Tr(\lasf_{a}\bro)$ \cite{bdqm02}. With this probability
distribution, we calculate entropic functions or other quantities
of interest. By $R_{\alpha}(\am;\bro)$ and $H_{\alpha}(\am;\bro)$,
we will mean the entropies (\ref{rpdf}) and (\ref{tsent})
calculated with $\Tr(\lasf_{a}\bro)$. Suppose that we further
measure another observable $\bn$. In the scenarios of successive
measurements, subsequent measurements are assumed to be performed
with a new ensemble of states. This formulation quite differs from
the tradition of preparation uncertainty relations.

Scenarios with successive measurement are formulated with respect
to the chosen form of post-first-measurement states
\cite{bfs2014}. In the first scenario, the second measurement is
performed on the state generated after the first measurement with
completely erased information. In terms of the projectors
$\lasf_{a}$, the pre-measurement state of the second measurement
is expressed as \cite{mdsrin03}
\begin{equation}
\Phi_{\am}(\bro)=\sum\nolimits_{a} \lasf_{a}\bro\,\lasf_{a}
\ . \label{fsmpm}
\end{equation}
To quantify the amount of uncertainty in successive
measurements, we will use quantities of the form
\begin{equation}
 R_{\alpha}(\am;\bro)+ R_{\beta}\bigl(\bn;\Phi_{\am}(\bro)\bigr)
\> , \label{rescp}
\end{equation}
and similarly with the corresponding Tsallis entropies.

The second scenario of successive measurements assumes that the
result of the first measurement should be kept. A focus on actual
measurement outcomes is typical for the so-called  selective
measurements. For example, incoherent selective measurements are
used in the formulation of monotonicity of coherence measures
\cite{bcp14}. Note that coherence quantifiers can be defined with
entropic functions of the Tsallis type \cite{rast16a}.

Thus, the second measurement is performed on the
post-first-measurement state conditioned on the actual measurement
outcome \cite{bfs2014,zzhang14}. It is represented as
$\vbro_{a}=r_{a}^{-1}\lasf_{a}\bro\,\lasf_{a}$, where
$r_{a}=\Tr(\lasf_{a}\bro)$. Measuring the second observable $\bn$
in each $\vbro_{a}$, we deal with the corresponding entropy $
R_{\beta}(\bn;\vbro_{a})$. Taking the average over all $a$, we
introduce the quantity
\begin{align}
&\sum\nolimits_{a} r_{a}\, R_{\beta}(\bn;\vbro_{a})=
\nonumber\\
&\sum\nolimits_{a} r_{a}\, R_{\alpha}(\am;\vbro_{a})
+\sum\nolimits_{a} r_{a}\, R_{\beta}(\bn;\vbro_{a})
\ . \label{avar}
\end{align}

Note that the first sum in the right-hand side of (\ref{avar}) is
zero. Indeed, measuring the observable $\am$ in its eigenstate
leads to a deterministic probability distribution, whence $
R_{\alpha}(\am;\vbro_{a})=0$ for all $a$. Thus, in finite
dimensions the left-hand side of  (\ref{avar}) is sufficient. This
is not the case for observables with continuous spectra. We cannot
say about a state, in which the measurement of position gives
exactly one particular value. Such states can be neither measured
nor prepared. Instead, we have to deal with well localized states
of finite, even small, scale. So, the right-hand side of
(\ref{avar}) will be useful in formulating the second scenario to
canonically conjugate variables.

In a similar manner, we can rewrite (\ref{avar}) with the use of
Tsallis' entropies. For $\alpha=\beta=1$, the quantity
(\ref{avar}) becomes the Shannon entropy averaged over all $a$.
The authors of \cite{bfs2014} applied the latter as a measure of
uncertainties in successive measurements. Uncertainty relations
for successive projective measurements in terms of R\'{e}nyi's
entropies were examined in \cite{zzhang14}. Formally, the sums
involved in (\ref{avar}) are very similar to one of the existing
definitions of conditional R\'{e}nyi's entropy. Note that the
proper definition of generalized conditional entropies is an open
question. There are several more or less justified approaches
\cite{tma12}. The simplest of them just lead to expressions of the
form (\ref{avar}). Also, the two kinds of conditional Tsallis
entropy are known in the literature \cite{sf06,rastkyb}. The
second form gives the quantity
\begin{equation}
\sum\nolimits_{a} r_{a}\, H_{\beta}(\bn;\vbro_{a})
\ . \label{avat}
\end{equation}
The conditional entropies mentioned above were used in studying
trade-off relations for noise and disturbance in
finite dimensions \cite{nodis16}. More properties of generalized
conditional entropies are discussed in \cite{rastit}.

When we deal with unbounded operators, the formulation should be
changed. Any real measurement apparatus is inevitably of a finite
size. Devices with a finite extension need a finite amount of
energy. Therefore, we cannot ask for some state, in which the
measurement of an observable gives exactly one particular value.
Of course, eigentates of position and momentum are often used as a
very convenient mathematical tool. They are not elements of the
Hilbert space, but can be treated in the context of rigged Hilbert
spaces \cite{mbg02}. In practice, we may deal with narrow
distributions that are of finite but small width. Measuring or
preparing some state with the particular value $\zeta$ of
position, one has to be affected by some vicinity of $\zeta$.
Therefore, we should treat each concrete result only as an
estimation. With more details, measurements of coordinates of a
microparticle are considered in chapter II of \cite{blokh73}.

To formulate the case of finite-resolution
measurements, new operators will be used. Let $\ay$ be
an observable with $\spc(\ay)=\mathbb{R}$, and let $|y\rangle$'s
be the eigenkets normalized through Dirac's delta function. For
some ``acceptance function'' $\zeta\mapsto{f}(\zeta)$, we define
\cite{paban13}
\begin{equation}
\km(\zeta):=\int\xdif{y}\,f(\zeta-y)\,|y\rangle\langle{y}|
\ . \label{kmzy}
\end{equation}
If the function $f$ obeys the normalization condition
\begin{equation}
\int|f(\zeta)|^{2}\,\xdif\zeta=1
\ , \label{afnrm}
\end{equation}
then the operators (\ref{kmzy}) satisfy
\begin{equation}
\int\xdif\zeta\> \km(\zeta)^{\dagger}\km(\zeta)=\pen
\ . \label{csrl}
\end{equation}
The acceptance function characterizes a degree of resolution of
the used measurement apparatus. From the physical viewpoint, we
may also assume that this function is even. When it is
sufficiently narrow, repeated trials with the pre-measurement
state will lead to a good estimation of the true probability
density function
\begin{equation}
w_{\bro}(y)=\langle{y}|\bro|y\rangle
\ . \label{wyr0}
\end{equation}
The experiment actually results in other density function. This
fact is also essential in deriving uncertainty relations for
characteristic functions \cite{tasca16}, where aperture
transmittance functions have been taken into account. The
probability density function dealt with is given by
\begin{equation}
P_{\bro}(\zeta)=
\int|f(\zeta-y)|^{2}\,w_{\bro}(y)\,\xdif{y}
\ . \label{wtr0}
\end{equation}
Due to (\ref{wtr0}), the dispersion of measured variable will be
added by the quantity \cite{paban13}
\begin{equation}
\sigma_{f}^{2}=\int\zeta^{2}\,|f(\zeta)|^{2}\,\xdif\zeta
\ . \label{sigfy}
\end{equation}
For good acceptance functions, any actual distortion of statistics
will be small. One of physically natural forms of acceptance
functions is the Gauss function \cite{paban13}. A physically
natural assumption is that a behavior of any acceptance function is
qualitatively similar. Even if it is narrow, its
tails are both non-zero, whence the density (\ref{wtr0}) cannot
take zero values exactly.

We shall now reformulate the first scenario of successive
measurements in the finite-resolution case. As was mentioned in
\cite{paban13}, the formula (\ref{fsmpm}) is replaced
with
\begin{equation}
\Phi_{\ay}(\bro)=
\int\xdif\zeta\> \km(\zeta)\bro\,\km(\zeta)^{\dagger}
\ . \label{fsmpm1}
\end{equation}
Note that the state (\ref{fsmpm1}) leads to the same probability
density function (\ref{wyr0}). But the latter is not directly
observable. As a consequence, the state (\ref{fsmpm1}) generates
the same distribution (\ref{wtr0}). The post-first-measurement
state (\ref{fsmpm1}) is further put into a finite-resolution
apparatus for the measurement of another observable $\ax$. This
second measurement is described in a similar manner. Let
$|x\rangle$'s be the eigenkets normalized also through Dirac's
delta function. The corresponding ``acceptance function''
$\xi\mapsto{g}(\xi)$ leads to operators of the form
\begin{equation}
\elm(\xi):=\int\xdif{x}\,g(\xi-x)\,|x\rangle\langle{x}|
\ . \label{elmx}
\end{equation}
Again, we cannot exactly determine the distribution with respect
to $x$ in any measured state. The variable $\xi$ is treated as an
estimation, for which the probability density function is
expressed similarly to (\ref{wtr0}). In the first scenario, this
density function is calculated with the post-first-measurement
state (\ref{fsmpm1}).

To characterize an amount of uncertainty in the first scenario of
successive measurements, we will introduce an analog of
(\ref{rescp}). In both the measurements, obtained statistics
actually deals with some estimation parameter. Hence, actual
probability density function should be taken into account. Using
entropies of continuous distributions, we will deal with the
quantity
\begin{equation}
R_{\alpha}\bigl(P;\bro\bigr)+
R_{\beta}\bigl(Q;\Phi_{\ay}(\bro)\bigr)
\>. \label{charqr00}
\end{equation}
Here, the R\'{e}nyi entropies are calculated due to (\ref{recon})
by substituting the probability density functions
$P_{\bro}(\zeta)$ and
\begin{equation}
Q(\xi)=
\int|g(\xi-x)|^{2}\,\langle{x}|\Phi_{\ay}(\bro)|x\rangle\,\xdif{x}
\ . \label{Wtr1}
\end{equation}
Another convenient approach is to calculate entropies with
binning. For instance, sampling of the function (\ref{wtr0})
into bins between marks $\zeta_{j}^{\prime}$ leads to a discrete
probability distribution $\pq_{\varDelta}^{(\zeta^{\prime})}$. In
the second measurement, entropies can be taken with binning
between some marks $\xi_{k}^{\prime}$. This approach leads to the
characteristic quantity
\begin{equation}
R_{\alpha}\bigl(\pq_{\varDelta}^{(\zeta^{\prime})};\bro\bigr)+
R_{\beta}\bigl(\qp_{\varDelta}^{(\xi^{\prime})};\Phi_{\ay}(\bro)\bigr)
\>, \label{charqr}
\end{equation}
and similarly for the case of Tsallis entropies. Entropic
quantities of the form (\ref{charqr}) provide a natural measure of
uncertainty in successive measurements of operators with
continuous spectra.

In the second scenario of successive measurements, each actual
result of the first measurement is retained. Assuming
$P_{\bro}(\zeta)\neq0$ in the corresponding domain, we now consider the
normalized output state
\begin{equation}
\vbro(\zeta)=P_{\bro}(\zeta)^{-1}\, \km(\zeta)\bro\,\km(\zeta)^{\dagger}
\ . \label{varz}
\end{equation}
Each $\vbro(\zeta)$ describes one of possible pre-measurement
states in the second measurement. Similarly to (\ref{charqr00}),
we then consider
\begin{equation}
{}\int R_{\alpha}\bigl(P;\vbro(\zeta)\bigr) P_{\bro}(\zeta)\,\xdif{\zeta} +
\int R_{\beta}\bigl(Q;\vbro(\zeta)\bigr) P_{\bro}(\zeta)\,\xdif{\zeta}
\ . \label{avapr00}
\end{equation}
The entropies $R_{\alpha}\bigl(P;\vbro(\zeta)\bigr)$ and
$R_{\beta}\bigl(Q;\vbro(\zeta)\bigr)$ are obtained with the
probability density functions $P(\tilde{\zeta})$ and
$Q(\tilde{\xi})$ determined for the given $\vbro(\zeta)$. We
further take the sum of entropies with binning for each
$\vbro(\zeta)$ and then average it over all $\zeta$. So, one
introduces the quantity
\begin{equation}
\int R_{\alpha}\bigl(\pq_{\varDelta}^{(\zeta^{\prime})};\vbro(\zeta)\bigr) P_{\bro}(\zeta)\,\xdif{\zeta}+
\int R_{\beta}\bigl(\qp_{\varDelta}^{(\xi^{\prime})};\vbro(\zeta)\bigr) P_{\bro}(\zeta)\,\xdif{\zeta}
\ . \label{avapr}
\end{equation}
The expressions (\ref{avapr00}) and (\ref{avapr}) generalize
(\ref{avar}) to the case of observables with continuous spectra.
In opposite to observables with discrete spectra, any state
$\vbro(\zeta)$ is associated with some uncertainty, even small, in
the first variable. The first term in each of the sums
(\ref{avapr00}) and (\ref{avapr}) is used to quantify this
feature. It is for this reason that the right-hand side of
(\ref{avar}) be written with adding zero term
$R_{\alpha}(\am;\vbro_{a})$. Its integral analog is strictly
non-zero in each of the sums (\ref{avapr00}) and
(\ref{avapr}). Furthermore, we may expect here some trade-off with
the second integral. The Tsallis-entropy version of uncertainty
measure for the second scenario is written similarly.

\section{Entropic uncertainty relations for successive measurements}\label{sec4}

In the preparation scenario, entropic uncertainty relations for
general canonically conjugate operators were proved in
\cite{barper15}. Such operators are obtained within the
Pegg--Barnett formalism as was explained in Section \ref{sec2}.
Let strictly positive numbers $\alpha$ and $\beta$ obey
$1/\alpha+1/\beta=2$ and $\alpha>1>\beta$. For any reference state
$\bsg$, one has \cite{barper15}
\begin{align}
\|w_{\bsg}\|_{\alpha}
&\leq\left(\frac{1}{2\pi}\right)^{(1-\beta)/\beta}\|W_{\bsg}\|_{\beta}
\ , \label{wyinf}\\
\|W_{\bsg}\|_{\alpha}
&\leq\left(\frac{1}{2\pi}\right)^{(1-\beta)/\beta}\|w_{\bsg}\|_{\beta}
\ . \label{wxinf}
\end{align}
Here, the probability density function $w_{\bsg}(y)$ is given by
substituting $\bsg$ into (\ref{wyr0}), and
$W_{\bsg}(x)=\langle{x}|\bsg|x\rangle$. The inequalities
(\ref{wyinf}) and (\ref{wxinf}) were derived by taking the limit
$d\to\infty$ in relations between functionals of discrete
probability distributions. From the mathematical viewpoint, the
derivation resembles the way by which the Hausdorff--Young
inequality follows from Riesz's theorem (see, e.g., \S\,8.17
of \cite{hardy}). In practice, we will deal with the probability
density functions with respect to either $\zeta$ or $\xi$. The
next step is to rewrite (\ref{wyinf}) and (\ref{wxinf}) in terms of
actually measured distributions.

Here, we recall one result for integral mean values with a weight
function (see theorem 204 of the book \cite{hardy}). It is similar
to Jensen's inequality. Let the weight function $\lambda(y)$ be
normalized. If $\phi^{\prime\prime}(t)$ is positive for all $t$
between $\inf{w}(y)$ and $\sup{w}(y)$, then
\begin{equation}
\phi\!\left(
\int\lambda(y)\,w(y)\,\xdif{y}
\right)
\leq
\int\lambda(y)\,\phi\bigl(w(y)\bigr)\,\xdif{y}
\ . \label{thm204}
\end{equation}
This holds under a lot of technical conditions concerning
positivity and integrability of functions. In our case, these
conditions are all satisfied. For $\alpha>1$, the function
$t\mapsto{t}^{\alpha}$ has positive second derivative. Applying
(\ref{thm204}) with $\lambda(y)=|f(\zeta-y)|^{2}$, we then get
\begin{equation}
[P_{\bsg}(\zeta)]^{\alpha}\leq
\int|f(\zeta-y)|^{2}\,[w_{\bsg}(y)]^{\alpha}\xdif{y}
\ . \label{thm2041}
\end{equation}
Integrating this inequality over $\zeta$ and using (\ref{afnrm}),
one gets
\begin{equation}
\|P_{\bsg}\|_{\alpha}^{\alpha}\leq\|w_{\bsg}\|_{\alpha}^{\alpha}
\ . \label{thm2042}
\end{equation}
For $0<\beta<1$, the function $t\mapsto{t}^{\beta}$ has negative
second derivative. In a similar manner, for the probability
density functions $W_{\bsg}(x)$ and
\begin{equation}
Q_{\bsg}(\xi)=\int|g(\xi-x)|^{2}\,W_{\bsg}(x)\,\xdif{x}
\ , \label{wtr01}
\end{equation}
we write
$\|W_{\bsg}\|_{\beta}^{\beta}\leq\|Q_{\bsg}\|_{\beta}^{\beta}$.
Combining the last two inequalities with (\ref{wyinf}) and
(\ref{wxinf}) finally gives
\begin{align}
\|P_{\bsg}\|_{\alpha}
&\leq\left(\frac{1}{2\pi}\right)^{(1-\beta)/\beta}\|Q_{\bsg}\|_{\beta}
\ , \label{pqinf}\\
\|Q_{\bsg}\|_{\alpha}
&\leq\left(\frac{1}{2\pi}\right)^{(1-\beta)/\beta}\|P_{\bsg}\|_{\beta}
\ , \label{qpinf}
\end{align}
where $1/\alpha+1/\beta=2$ and $\alpha>1>\beta$. These relations
initially hold for the preparation scenario with arbitrary
reference state $\bsg$. However, they are still valid for two
different states that have the same probability density function
in one of the two measured observables.

Formulating entropic uncertainty relations for successive
measurements, we begin with the second scenario. Let
$\vbro(\zeta)$ be one of particular outputs of the first
measurement, in which we have measured $\ay$. Substituting
$\bsg=\vbro(\zeta)$ and taking the logarithm of (\ref{pqinf}) and
(\ref{qpinf}), we finally obtain
\begin{equation}
R_{\alpha}\bigl(P;\vbro(\zeta)\bigr)+
R_{\beta}\bigl(Q;\vbro(\zeta)\bigr)
\geq\ln2\pi
\ . \label{ressc010}
\end{equation}
We now average (\ref{ressc010}) with the probability density
function $P_{\bro}(\zeta)$ calculated for the input $\bro$. Under
the condition $1/\alpha+1/\beta=2$, one gets
\begin{align}
&\int R_{\alpha}\bigl(P;\vbro(\zeta)\bigr) P_{\bro}(\zeta)\,\xdif{\zeta}+
\int R_{\beta}\bigl(Q;\vbro(\zeta)\bigr) P_{\bro}(\zeta)\,\xdif{\zeta}
\nonumber\\
&\geq\ln2\pi
\ . \label{ressc100}
\end{align}
The relation differs from relations for finite-dimensional
observables in the following respect. Eigenstates of an observable
with continuous spectra are unphysical and cannot be prepared. For
each particular output of the first measurement, measuring the
same observable on $\vbro(\zeta)$ will lead to some distribution
of outcomes. A width of this distribution is always non-zero,
though small for a good acceptance function. Moreover, our
attempts to reduce values of the first integral will lead to
increasing lower bound on the second integral. This question does
not appear in successive measurements of observables with pure
point spectra. Indeed, the first sum in the right-hand side of
(\ref{avar}) is always zero.

Entropic uncertainty relations in the first scenario are based on
an additional observation. Let $\ay$ be the first of the two measured
observables. As was mentioned above, the first-measurement output
$\Phi_{\ay}(\bro)$ and the input state $\bro$ share the same
density function $w_{\bro}(y)$. Due to (\ref{wtr0}), they also
have the same distribution $P_{\bro}(\zeta)$, whence
\begin{equation}
R_{\alpha}\bigl(P;\bro\bigr)=R_{\alpha}\bigl(P;\Phi_{\ay}(\bro)\bigr)
\, , \label{eqenrr0}
\end{equation}
Rewriting (\ref{ressc010}) with $\Phi_{\ay}(\bro)$ instead of
$\vbro(\zeta)$, we obtain uncertainty relations in the first
scenario of successive measurements due to (\ref{eqenrr0}). If
positive numbers $\alpha$ and $\beta$ obey $1/\alpha+1/\beta=2$,
then
\begin{equation}
R_{\alpha}\bigl(P;\bro\bigr)+
R_{\beta}\bigl(Q;\Phi_{\ay}(\bro)\bigr)
\geq\ln2\pi
\ . \label{ressc000}
\end{equation}
Thus, we formulated uncertainty relations for successive
measurements of canonically conjugate variables in terms of
R\'{e}nyi's entropies of continuous distributions.

To obtain definitely positive quantities, we can use entropies
with binning. The only step is to convert (\ref{pqinf}) and
(\ref{qpinf}) into relations between the corresponding discrete
distributions. Instead of $P_{\bsg}(\zeta)$, we now deal with the
probabilities
\begin{equation}
p_{j}^{(\zeta^{\prime})}:=\int\nolimits_{\zeta_{j}^{\prime}}^{\zeta_{j+1}^{\prime}}
P_{\bsg}(\zeta){\,}\xdif\zeta
\ , \label{pjpz}
\end{equation}
which form the distribution $\pq_{\varDelta}^{(\zeta^{\prime})}$.
The basic size of discretization is
$\Delta\zeta=\max\Delta\zeta_{j}^{\prime}$, where
$\Delta\zeta_{j}^{\prime}=\zeta_{j+1}^{\prime}-\zeta_{j}^{\prime}$.
For the second probability density function, we introduce the
probabilities
\begin{equation}
q_{k}^{(\xi^{\prime})}:=\int\nolimits_{\xi_{k}^{\prime}}^{\xi_{k+1}^{\prime}}
Q_{\bsg}(\xi){\,}\xdif\xi
\ . \label{qkqx}
\end{equation}
They form the second probability distribution
$\qp_{\varDelta}^{(\xi^{\prime})}$. We also put the intervals
$\Delta\xi_{k}^{\prime}=\xi_{k+1}^{\prime}-\xi_{k}^{\prime}$ and
their maximum $\Delta\xi=\max\Delta\xi_{k}^{\prime}$. Combining
(\ref{pqinf}) and (\ref{qpinf}) with the corresponding results of
the form (\ref{cnv1}) and (\ref{cnv2}), the following conclusion
takes place. Under the conditions $1/\alpha+1/\beta=2$ and
$\alpha>1>\beta$, we have
\begin{align}
\bigl\|\pq_{\varDelta}^{(\zeta^{\prime})}\bigr\|_{\alpha}
&\leq\left(\frac{\Delta\zeta\Delta\xi}{2\pi}\right)^{(1-\beta)/\beta}\bigl\|\qp_{\varDelta}^{(\xi^{\prime})}\bigr\|_{\beta}
\ , \label{pqinfz}\\
\bigl\|\qp_{\varDelta}^{(\xi^{\prime})}\bigr\|_{\alpha}
&\leq\left(\frac{\Delta\zeta\Delta\xi}{2\pi}\right)^{(1-\beta)/\beta}\bigl\|\pq_{\varDelta}^{(\zeta^{\prime})}\bigr\|_{\beta}
\ . \label{qpinfx}
\end{align}
Entropic uncertainty relations with binning are derived from
(\ref{pqinfz}) and (\ref{qpinfx}) similarly to the way by which
the results (\ref{ressc100}) and (\ref{ressc000}) follow from
(\ref{pqinf}) and (\ref{qpinf}). In terms of R\'{e}nyi entropies,
we have
\begin{align}
& R_{\alpha}\bigl(\pq_{\varDelta}^{(\zeta^{\prime})};\bro\bigr)+
 R_{\beta}\bigl(\qp_{\varDelta}^{(\xi^{\prime})};\Phi_{\ay}(\bro)\bigr)
\geq\ln\!\left(\frac{2\pi}{\Delta\zeta\Delta\xi}\right)
, \label{ressc00}\\
& \int R_{\alpha}\bigl(\pq_{\varDelta}^{(\zeta^{\prime})};\vbro(\zeta)\bigr) P_{\bro}(\zeta)\,\xdif{\zeta}+
\int R_{\beta}\bigl(\qp_{\varDelta}^{(\xi^{\prime})};\vbro(\zeta)\bigr) P_{\bro}(\zeta)\,\xdif{\zeta}
\nonumber\\
& \geq\ln\!\left(\frac{2\pi}{\Delta\zeta\Delta\xi}\right)
, \label{ressc1}
\end{align}
where $1/\alpha+1/\beta=2$. Non-zero lower bounds imply a trade-off
between values of the first and second entropic measures of
uncertainties in each of the formulas (\ref{ressc00}) and
(\ref{ressc1}). To ensure positive bounds, the interval
characteristics $\Delta\zeta$ and $\Delta\xi$ should be chosen
small enough. It is connected with the fact that one quantum
degree of freedom occupies a dimensionless phase cell with a size
not less than $2\pi$. As was already noted right after
(\ref{cnv1}) and (\ref{cnv2}), entropies with binning becomes
generally unbounded in the limit of zero bins. In this limit, the
inequalities (\ref{ressc00}) and (\ref{ressc1}) do not give an
informative statement. Instead, we can use here (\ref{ressc100})
and (\ref{ressc000}).

To get uncertainty relations in terms of R\'{e}nyi entropies,
purely algebraic operations were used. The case of Tsallis
entropies is not so immediate. Our approach is based on the method
of \cite{rast104}, where the minimization problem was examined. We
refrain from presenting the details here. The results are
formulated as follows. Let strictly positive numbers $\alpha$ and
$\beta$ obey $1/\alpha+1/\beta=2$. For any input state $\bro$, we
have
\begin{align}
& H_{\alpha}\bigl(\pq_{\varDelta}^{(\zeta^{\prime})};\bro\bigr)+
 H_{\beta}\bigl(\qp_{\varDelta}^{(\xi^{\prime})};\Phi_{\ay}(\bro)\bigr)
\geq\ln_{\mu}\!\left(\frac{2\pi}{\Delta\zeta\Delta\xi}\right)
, \label{tessc00}\\
& \int H_{\alpha}\bigl(\pq_{\varDelta}^{(\zeta^{\prime})};\vbro(\zeta)\bigr) P_{\bro}(\zeta)\,\xdif{\zeta}+
\int H_{\beta}\bigl(\qp_{\varDelta}^{(\xi^{\prime})};\vbro(\zeta)\bigr) P_{\bro}(\zeta)\,\xdif{\zeta}
\nonumber\\
& \geq\ln_{\mu}\!\left(\frac{2\pi}{\Delta\zeta\Delta\xi}\right)
, \label{tessc1}
\end{align}
where $\mu=\max\{\alpha,\beta\}$. Hence, uncertainty relations in
both the scenarios of successive measurements of canonically
conjugate operators are expressed in terms of Tsallis entropies.
These entropic uncertainty relations also show complementarity in
successive measurements of general canonically conjugate
operators.

We finally consider the position and momentum operators, which
give a primary case of the canonical commutation relation. Here,
the above entropic uncertainty relations can be improved. In
general, the Riesz theorem {\it per se} provides only an upper
bound on the corresponding norm of a linear transformation.
Calculating exact value of the required norm may lead to improved
relations. For the Fourier transform, the exact value was found by
Beckner \cite{beck}. As was shown by Bia{\l}ynicki-Birula and
Mycielski \cite{birula1}, this result implies an improvement of
Hirschman's uncertainty relation \cite{hirs}. The wave functions
in the position and momentum spaces are connected via the Fourier
transform. Using the Beckner result, we replace $2\pi$ with
$\varkappa\pi$ in the formulas (\ref{wyinf}) and (\ref{wxinf}).
The square of $\varkappa$ is expressed as \cite{barper15}
\begin{equation}
\varkappa^{2}=\alpha^{1/(\alpha-1)}\beta^{1/(\beta-1)}
\, . \label{mpmpf}
\end{equation}
The parameter $\varkappa$ grows from $\varkappa=2$ for
$\beta=1/2$ up to $\varkappa=e$ for $\beta=1$. In the first
scenario, we finally obtain
\begin{align}
& R_{\alpha}\bigl(P;\bro\bigr)+
 R_{\beta}\bigl(Q;\Phi_{\ay}(\bro)\bigr)
\geq\ln\varkappa\pi
\ , \label{ressccon}\\
& R_{\alpha}\bigl(\pq_{\varDelta}^{(\zeta^{\prime})};\bro\bigr)+
 R_{\beta}\bigl(\qp_{\varDelta}^{(\xi^{\prime})};\Phi_{\ay}(\bro)\bigr)
\geq\ln\!\left(\frac{\varkappa\pi}{\Delta\zeta\Delta\xi}\right)
, \label{resscpm}
\end{align}
where $1/\alpha+1/\beta=2$. In a similar manner, we recast the
uncertainty relations (\ref{ressc100}) and (\ref{ressc1})
concerning the second scenario of successive measurements. In
terms of the Tsallis entropies, we merely rewrite the right-hand
sides of (\ref{tessc00}) and (\ref{tessc1}) with $\varkappa\pi$
instead of $2\pi$. In the case $\alpha=\beta=1$, all the above
relations are expressed via the Shannon entropies. In the lower
bounds, the term $\varkappa\pi$ then becomes $e\pi$. The
right-hand side of (\ref{resscpm}) was derived as entropic bounds
in \cite{barper15}. However, that paper concerns the traditional
formulation of two independent measurements in the same
pre-measurement state. Such lower bounds remain sometimes valid
for the scenarios of successive measurements.

\section{Conclusions}\label{sec5}

We have formulated entropic uncertainty relations for successive
measurements of general canonically conjugate observables. As
canonical conjugacy plays a crucial role in physics, all its
aspects deserve investigations. Studies of scenarios with
successive measurements are essential from several viewpoints. As
was noticed in sect. 5.5 of \cite{bdqm02}, the concept of wave
function reduction becomes of interest only if one performs at
least two successive measurements on a system. The most
traditional scenario is known as preparation uncertainty
relations. However, this picture seems to be too idealized. It
differs from the Heisenberg's thought experiment
\cite{heisenberg}, whence uncertainty relations are all
originated. Also, the considered successive measurements are not
projective. It was shown that entropic measures of uncertainty for
such measurements should be introduced in a manner different from
the case of finite-dimensional observables. One of distinctions
concerns a proper form of the state right after the first
measurement. The post-first-measurement state was chosen according
to the two possible scenarios.

Successive measurements of observables with continuous spectra
depend on acceptance functions of the apparatuses. It was shown
how to treat such measurements within entropic approach. The
entropic uncertainty relations derived explicitly reveal
complementarity in successive measurements of canonically
conjugate operators. The presented results give evidence that
entropic uncertainty bounds of the traditional formulation may
hold in more realistic situations. We examined the case when the
conjugate observables both have continuous spectra. Similar ideas
can be applied to such pairs as the angle-angular momentum pair
and the optical phase-number pair. The latter may need an
additional consideration due to a construction of the
Hermitian phase operator. We hope to address this case in future
investigations. One of open questions is to get
information-theoretic formulation of noise-disturbance relations
for canonically conjugate variables. In finite dimensions, such
approaches were proposed in \cite{bhow13,cofu13}.

\acknowledgments

The author is grateful to anonymous referees for valuable comments.


\begin{thebibliography}{100}

\bibitem{heisenberg}%---------------------------------------------
W.~Heisenberg, Z. Phys. {\bf 43}, 172 (1927).

\bibitem{kennard}%------------------------------------------------
E.\,H.~Kennard, Z. Phys. {\bf 44}, 326 (1927).

\bibitem{robert}%--------------------------------------------------
H.\,P.~Robertson, Phys. Rev. {\bf 34}, 163 (1929).

\bibitem{deutsch}%---------------------------------------------------
D.~Deutsch, Phys. Rev. Lett. {\bf 50}, 631 (1983).

\bibitem{maass}%----------------------------------------------------
H.~Maassen and J.\,B.\,M.~Uffink, Phys. Rev. Lett. {\bf 60}, 1103 (1988).

\bibitem{hirs}%-------------------------------------------------
I.\,I.~Hirschman, Amer. J. Math. {\bf 79}, 152 (1957).

\bibitem{beck}%-------------------------------------------------
W.~Beckner, Ann. Math. {\bf 102}, 159 (1975).

\bibitem{birula1}%-------------------------------------------------
I.~Bia{\l}ynicki-Birula and J.~Mycielski, Commun. Math. Phys. {\bf 44}, 129 (1975).

\bibitem{walborn09}%------------------------------------------------
S.\,P.~Walborn, B.\,G.~Taketani, A.~Salles, F.~Toscano, and R.\,L.~de Matos Filho, Phys. Rev. Lett. {\bf 103}, 050401 (2009).

\bibitem{huang11}%------------------------------------------------
Y.~Huang, Phys. Rev. A {\bf 83}, 052124 (2011).

\bibitem{ww10}%-------------------------------------------------
S.~Wehner and A.~Winter, New J. Phys. {\bf 12}, 025009 (2010).

\bibitem{brud11}%------------------------------------------------
I.~Bia{\l}ynicki-Birula and {\L}.~Rudnicki, in: Statistical
Complexity, edited by K.\,D.~Sen (Springer, Berlin, 2011), 1--34.

\bibitem{cbtw15}%------------------------------------------------
P.\,J.~Coles, M.~Berta, M.~Tomamichel, and S.~Wehner, E-print arXiv:1511.04857 [quant-ph] (2015).

\bibitem{asmf15}%------------------------------------------------
K.~Abdelkhalek, R.~Schwonnek, H.~Maassen, F.~Furrer, J.~Duhme,
P.~Raynal, B.-G.~Englert, and R.\,F.~Werner, E-print
arXiv:1509.00398 [quant-ph] (2015).

\bibitem{huang12}%------------------------------------------------
Y.~Huang, Phys. Rev. A {\bf 86}, 024101 (2012).

\bibitem{mpati14}%--------------------------------------------------
L.~Maccone and A.\,K.~Pati, Phys. Rev. Lett. {\bf 113}, 260401 (2014).

\bibitem{prz13}%------------------------------------------------
Z.~Pucha{\l}a, {\L}.~Rudnicki, and K.~\.{Z}yczkowski, J. Phys. A: Math. Theor. {\bf 46},
272002 (2013).

\bibitem{fgg13}%------------------------------------------------------------
S.~Friedland, V.~Gheorghiu, and G.~Gour, Phys. Rev. Lett. {\bf 111}, 230401 (2013).

\bibitem{rpz14}%------------------------------------------------
{\L}.~Rudnicki, Z.~Pucha{\l}a, and K.~\.{Z}yczkowski, Phys. Rev. A {\bf 89}, 052115 (2014).

\bibitem{lrud15}%------------------------------------------------
{\L}.~Rudnicki, Phys. Rev. A {\bf 91}, 032123 (2015).

\bibitem{arkz16}%------------------------------------------------
A.\,E.~Rastegin and K.~\.{Z}yczkowski, J. Phys. A: Math. Theor. {\bf 49}, 355301 (2016).

\bibitem{msp16}%------------------------------------------------
C.~Mukhopadhyay, N.~Shukla, and A.\,K.~Pati, Europhys. Lett. {\bf 113}, 50002 (2015).

\bibitem{gour16}%------------------------------------------------
V.~Narasimhachar, A.~Poostindouz, and G.~Gour, New J. Phys. {\bf 18}, 033019 (2016).

\bibitem{mdsrin03}%------------------------------------------------
M.\,D.~Srinivas, Pramana J. Phys. {\bf 60}, 1137 (2003).

\bibitem{paban13}%--------------------------------------------------
J.~Distler and S.~Paban, Phys. Rev. A {\bf 87}, 062112 (2013).

\bibitem{renner10}%------------------------------------------------
M.~Berta, M.~Christandl, R.~Colbeck, J.\,M.~Renes and R.~Renner, Nature Phys. {\bf 6}, 659 (2010).

\bibitem{renner11}%------------------------------------------------
M.~Tomamichel and R.~Renner, Phys. Rev. Lett. {\bf 106}, 110506 (2011).

\bibitem{nbw12}%------------------------------------------------
N.\,H.\,Y.~Ng, M.~Berta, and S.~Wehner, Phys. Rev. A {\bf 86}, 042315 (2012).

\bibitem{wineland13}%------------------------------------------------
D.~Wineland, Ann. Phys. (Berlin) {\bf 525}, 739 (2013).

\bibitem{haroche13}%------------------------------------------------
S.~Haroche, Ann. Phys. (Berlin) {\bf 525}, 753 (2013).

\bibitem{blw13}%------------------------------------------------
P.~Busch, P.~Lahti, and R.\,F.~Werner, Phys. Rev. Lett. {\bf 111}, 160405 (2013).

\bibitem{ozawa03}%--------------------------------------------------
M.~Ozawa, Phys. Rev. A {\bf 67}, 042105 (2003).

\bibitem{ozawa04}%--------------------------------------------------
M.~Ozawa, Ann. Phys. {\bf 311}, 350 (2004).

\bibitem{bfs2014}%-------------------------------------------------
K.~Baek, T.~Farrow, and W.~Son, Phys. Rev. A {\bf 89}, 032108 (2014).

\bibitem{zzhang14}%------------------------------------------------
J.~Zhang, Y.~Zhang, and C.-S.~Yu, Quantum Inf. Process. {\bf 14}, 2239 (2015).

\bibitem{rastctp15}%------------------------------------------------
A.\,E.~Rastegin, Commun. Theor. Phys. {\bf 63}, 687 (2015).

\bibitem{bengtsson}%------------------------------------------------
I.~Bengtsson and K.~\.{Z}yczkowski, Geometry of Quantum States: An
Introduction to Quantum Entanglement (Cambridge University Press,
Cambridge, 2006).

\bibitem{renyi61}%-------------------------------------------------
A.~R\'{e}nyi, in: Proceedings of the 4th Berkeley Symposium on
Mathematical Statistics and Probability, edited by J.~Neyman (University of California Press, Berkeley, 1961), 547--561.

\bibitem{jizba}%-------------------------------------------------
P.~Jizba and T.~Arimitsu, Ann. Phys. {\bf 312}, 17 (2004)

\bibitem{tsallis}%------------------------------------------------
C.~Tsallis, J. Stat. Phys. {\bf 52}, 479 (1988).

\bibitem{havrda}%-----------------------------------------------
J.~Havrda and F.~Charv\'{a}t, Kybernetika {\bf 3}, 30 (1967).

\bibitem{rastann}%---------------------------------------------------------------
A.\,E.~Rastegin, Ann. Phys. {\bf 355}, 241 (2015).

\bibitem{ecount}%---------------------------------------------------------------
A.\,E.~Rastegin, Graphs Combin., DOI:10.1007/s00373-016-1731-x.

\bibitem{IBB84}%------------------------------------------------
I.~Bia{\l}ynicki-Birula, Phys. Lett. A {\bf 103}, 253 (1984).

\bibitem{IBB06}%------------------------------------------------
I.~Bia{\l}ynicki-Birula, Phys. Rev. A {\bf 74}, 052101 (2006).

\bibitem{taska13}%------------------------------------------------
D.\,S.~Tasca, {\L}.~Rudnicki, R.\,M.~Gomes, F.~Toscano, and S.\,P.~Walborn, Phys. Rev. Lett. {\bf 110}, 210502 (2013).

\bibitem{dixon13}%------------------------------------------------
J.~Schneeloch, P.\,B.~Dixon, G.\,A.~Howland, C.\,J.~Broadbent, and J.\,C.~Howell, Phys. Rev. Lett. {\bf 110}, 130407 (2013).

\bibitem{howell14}%------------------------------------------------
J.~Schneeloch, C.\,J.~Broadbent, and J.\,C.~Howell, Phys. Lett. A {\bf 378}, 766 (2014).

\bibitem{hardy}%----------------------------------------------------
G.\,H.~Hardy, J.\,E.~Littlewood, and G.~Polya, Inequalities (Cambridge University Press, London, 1934).

\bibitem{barper15}%------------------------------------------------
A.\,E.~Rastegin, Found. Phys. {\bf 45}, 923 (2015).

\bibitem{PBepl}%---------------------------------------------------------
D.\,T.~Pegg and S.\,M.~Barnett, Europhys. Lett. {\bf 6}, 483 (1988).

\bibitem{BP89}%--------------------------------------------------
S.\,M.~Barnett and D.\,T.~Pegg, J. Mod. Opt. {\bf 36}, 7 (1989).

\bibitem{PB89}%--------------------------------------------------
D.\,T.~Pegg and S.\,M.~Barnett, Phys. Rev. A {\bf 39}, 1665 (1989).

\bibitem{dirac27}%---------------------------------------------------------
P.\,A.\,M.~Dirac, Proc. R. Soc. A {\bf 114}, 243 (1927).

\bibitem{lynch}%--------------------------------------------------
R.~Lynch, Phys. Rep. {\bf 256}, 367 (1995).

\bibitem{CN68}%------------------------------------------------
P.~Carruthers and M.\,M.~Nieto, Rev. Mod. Phys. {\bf 40}, 411 (1968).

\bibitem{pvb90}%----------------------------------------------------
D.\,T.~Pegg, J.\,A.~Vaccaro, and S.\,M.~Barnett, J. Mod. Opt. {\bf 37}, 1703 (1990).

\bibitem{gvb95}%----------------------------------------------------
A.\,R.~Gonzalez, J.\,A.~Vaccaro, and S.\,M.~Barnett, Phys. Lett. A {\bf 205}, 247 (1995).

\bibitem{schwinger60}%-------------------------------------------------
J.~Schwinger, Proc. Natl. Acad. Sci. {\bf 46}, 570 (1960).

\bibitem{riesz27}%---------------------------------------------------
M.~Riesz, Acta Math. {\bf 49}, 465 (1927).

\bibitem{bdqm02}%------------------------------------------------
J.-L.~Basdevant and J.~Dalibard, Quantum Mechanics (Springer, Berlin, 2002).

\bibitem{bcp14}%-----------------------------------------------
T.~Baumgratz, M.~Cramer, and M.\,B.~Plenio, Phys. Rev. Lett. {\bf 113}, 140401 (2014).

\bibitem{rast16a}%-----------------------------------------------
A.\,E.~Rastegin, Phys. Rev. A {\bf 93}, 032136 (2016).

\bibitem{tma12}%-------------------------------------------------
A.~Teixeira, A.~Matos, and L.~Antunes, IEEE Trans. Inf. Theory {\bf 58}, 4273 (2012).

\bibitem{sf06}%-------------------------------------------------
S.~Furuichi, J. Math. Phys. {\bf 47}, 023302 (2006).

\bibitem{rastkyb}%------------------------------------------------
A.\,E.~Rastegin, Kybernetika {\bf 48}, 242 (2012).

\bibitem{nodis16}%------------------------------------------------
A.\,E.~Rastegin, Quantum Inf. Comput. {\bf 16}, 0313 (2016).

\bibitem{rastit}%-----------------------------------------------
A.\,E.~Rastegin, RAIRO--Theor. Inf. Appl. {\bf 49}, 67 (2015).

\bibitem{mbg02}%-------------------------------------------------
R.~de la Madrid, A.~Bohm, and M.~Gadella, Fortschr. Phys. {\bf 50}, 185 (2002).

\bibitem{blokh73}%------------------------------------------------
D.\,I.~Blokhintsev, Space and Time in the Microworld (D. Reidel Publishing Company, Dordrecht, 1973).

\bibitem{tasca16}%------------------------------------------------
{\L}.~Rudnicki, D.\,S.~Tasca, and S.\,P.~Walborn, Phys. Rev. A {\bf 93}, 022109 (2016).

\bibitem{rast104}%----------------------------------------------------
A.\,E.~Rastegin, J. Phys. A: Math. Theor. {\bf 44}, 095303 (2011).

\bibitem{bhow13}%------------------------------------------------
F.~Buscemi, M.\,J.\,W.~Hall, M.~Ozawa, and M.\,M.~Wilde, Phys. Rev. Lett. {\bf 112}, 050401 (2014).

\bibitem{cofu13}%------------------------------------------------
P.\,J.~Coles and F.~Furrer, Phys. Lett. A {\bf 379}, 105 (2015).

\end{thebibliography}
\end{document}